\documentclass[aps,pra,twocolumn,floatfix,superscriptaddress,showpacs]{revtex4-1}
\usepackage{graphicx}
\usepackage{amssymb,amsmath}
\usepackage{color}

\newcommand{\ket}[1]{|#1\rangle}				
\newcommand{\bra}[1]{\langle #1|}				

\begin{document}
\title{A hybrid model for Rydberg gases including exact two-body correlations}

\author{Kilian P. Heeg}
\affiliation{Max-Planck-Institut f\"ur Kernphysik, Saupfercheckweg 1, 69117 Heidelberg, Deutschland}
\author{Martin G\"arttner}
\affiliation{Max-Planck-Institut f\"ur Kernphysik, Saupfercheckweg 1, 69117 Heidelberg, Deutschland}
\affiliation{Institut f\"ur Theoretische Physik, Ruprecht-Karls-Universit\"at Heidelberg, Philosophenweg~16, 69120~Heidelberg, Germany}
\author{J\"org Evers}
\affiliation{Max-Planck-Institut f\"ur Kernphysik, Saupfercheckweg 1, 69117 Heidelberg, Deutschland}

\date{\today}
\begin{abstract}
	A model for the simulation of ensembles of laser-driven Rydberg-Rydberg interacting multi-level atoms is discussed. Our hybrid approach combines an exact two-body treatment of nearby atom pairs with an effective approximate treatment for spatially separated pairs. We propose an optimized evolution equation based only on the system steady state, and a time-independent Monte Carlo technique is used to efficiently determine this steady state. The hybrid model predicts features in the pair correlation function arising from multi-atom processes which existing models can only partially reproduce. Our interpretation of these features shows that higher-order correlations are relevant already at low densities. Finally, we analyze the performance of our model in the high-density case.
\end{abstract}
\maketitle

\section{Introduction}
The distinctive properties of Rydberg atoms~\cite{Saffman10,Comparat10,Loew12} render them a powerful implementation of a tunable strongly interacting quantum many-body system. The signature effect of Rydberg atoms is the dipole blockade~\cite{Lukin01,Jaksch00,Singer04,Tong04,Urban09,Gaetan09}, which can be understood using rather basic theoretical models~\cite{Tong04,Chotia08,Weimer08}. But ongoing experimental progress provides means to explore the correlations on a deeper level. Accordingly, more involved observables such as the Mandel $Q$ parameter \cite{cubellie05,reinh08}, the pair correlation function~\cite{guenth12,Olmos11,Schwarzkopf11} or quantum optical effects in the presence of Rydberg interactions~\cite{Schempp10,Pritchard10,Dudin12} moved into the focus of interest. 
State-of-the-art experiments push existing models for Rydberg gases to the limits of their validity ranges \cite{guenth12,Schwarzkopf11}, and require exceedingly long simulation times. Therefore, better simulation techniques are highly desirable.

One approach to simulate the many-body system is to truncate the otherwise exponentially growing state space, e.g., based on the Rydberg blockade. Such calculations of the exact Hamiltonian dynamics in a truncated Hilbert space provide valuable insights \cite{Weimer08,Rob05,Younge09,Olmos09,Gaerttner12,Gaerttner12b}. However, they are not well suited to model weakly interacting or low-density gases, as then the state space becomes too large. Due to the treatment on the level of the wave function, incoherent effects like spontaneous emission or dephasing cannot be incorporated in a straightforward way. Also, two-step excitation schemes cannot be described, as the intermediate state would spoil the state space reduction. But especially for EIT experiments \cite{Pritchard10,EIT} a three-level picture is crucial. In this case dedicated theories for the light and its properties have been developed \cite{Petrosyan11,Sevincli11,Peyronel12}, however at the cost of an accurate atomic description.

An alternative approach to model the many-atom system is to approximate the inter-atomic correlations. 
The simplest approach is the mean field approximation $\langle A_i B_j \rangle\approx\langle A_i\rangle \langle B_j \rangle$ for operators $A,B$ acting on two different atoms $i,j$, which already describes the Rydberg blockade well \cite{Tong04,Chotia08}. However, it fails for higher correlations and for more involved observables~\cite{Schempp10,Pritchard10}. 
A cluster expansion to higher-order atomic correlations could successfully describe an experiment on coherent population trapping in Rydberg atoms~\cite{Schempp10}, but this method turned out to give inconsistent results at high densities \cite{Ates11}. 
Alternatively, a rate equation model was introduced in \cite{Ates07}. Using an adiabatic elimination of the atomic coherences, an effective rate equation for the atomic populations alone could be derived. The resulting equation system still grows exponentially in the number of atoms, but can be solved using Monte Carlo techniques. This method is capable of modelling a multi-step excitation~\cite{Ates11}. However, due to the simplified treatment of the interaction, only single-atom processes can be described and correlations are approximated.

\begin{figure}[b]
	\centering
	\includegraphics{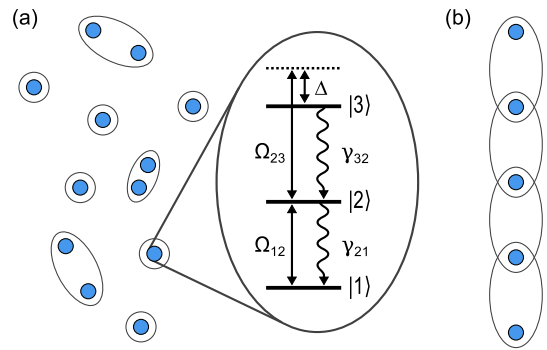}
	\caption{(Color online) \textbf{(a)}~Our model divides atoms into pairs and single atoms allowing an exact treatment of the two-body interaction up to a inter-atomic separation $L_{Cr}$. Encircled subsets interact with each other via an effective detuning. \textbf{(b)}~On a lattice structure, overlapping pairs have to be used.}
	\label{fig:level_scheme}
\end{figure}

In this work, we discuss a simulation technique combining higher predictive power with fast calculation times. In our model, atom pairs with distances below a characteristic length scale $L_{Cr}$ are treated exactly, resulting in accurate two-body correlations. Atoms with distances above $L_{Cr}$ are incorporated via effective detunings. We show that this hybrid approach describes three-body systems well over the whole range of interaction strengths. To efficiently determine the many-body steady state, we eliminate the time-consuming calculations of transition rates in each step based on a time-independent Monte Carlo technique. We show the consistency of our approach with the relevant existing models for simple observables as the excitation fraction of a disordered Rydberg ensemble. In contrast, the hybrid model predicts structures in the pair correlation function, which the existing models cannot fully reproduce. The same resonances can be seen in the excitation statistics. These structures arise from 
higher-order processes, which we show to be relevant already at low densities. Finally, we discuss how our ansatz can be expanded to cover parameter ranges of high densities. This opens perspectives for the highly desirable modelling of extended clouds of multi-level atoms at high densities, which is not possible with numerical integrations of Schr\"odinger's equation on truncated Hilbert spaces.

\section{Model and Methods}
\subsection{Motivation}
\begin{figure}
	\centering
	\includegraphics{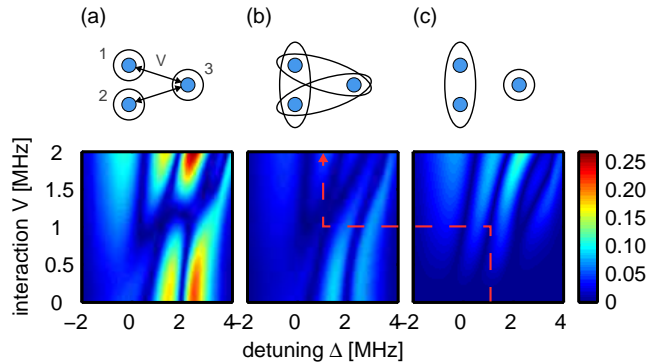}
	\caption{(Color online) Three models consisting of lower order subsets are investigated as simplified models for the $N=3$ case. The relative deviation $|\rho_{33}^{\text{simpl.}} - \rho_{33}^{\text{exact}}|/\rho_{33}^{\text{exact}}$ from the exact solution is shown for varying detuning $\Delta$ and interaction $V$ with the third atom. Models including the exact treatment of pairs give more precise results than the rate equation. The dashed line depicts the models which are suited best for a given $V$. Parameters: $\Omega_{12} = 3~\text{MHz}, \Omega_{23}=2~\text{MHz},\gamma_{21}=6~\text{MHz},\gamma_{32}=25~\text{kHz}, V_{12}=2~\text{MHz}$.}
	\label{fig:N3deviation}
\end{figure}

We consider a frozen gas~\cite{Comparat10} of three-level atoms driven by a two-step excitation scheme as shown in Fig.~\ref{fig:level_scheme}. The lower transition between ground state $|1\rangle$ and intermediate state $|2\rangle$ is driven resonantly with Rabi frequency $\Omega_{12}$, and the upper transition between intermediate state $|2\rangle$ and Rydberg state $|3\rangle$ is driven with detuning $\Delta = \omega - \omega_{32}$ and Rabi frequency $\Omega_{23}$. We note that generally also different level schemes, such as a two-level system or a true two-photon excitation with strongly detuned intermediate level \cite{Cardimona93} can be simulated.

Before we start with the theoretical analysis, we first illustrate the main idea. We start from a two-atom master equation which yields exact results for two atoms.
But there is no unique method to extend this ansatz for pairs to many atoms without using larger atom subsets. To illustrate this, we consider a system of three atoms in which the first two atoms interact strongly with coupling $V_{12}$. A third atom is moved from large distance ($V:=V_{13}=V_{23}=0$) towards the pair until the three atoms form an equilateral triangle. 
For the resulting interaction energies $V$, we compare the exact three-body steady state solution with different simplified models consisting of single atoms or pairs. The result for different detunings $\Delta$ is shown in Fig.~\ref{fig:N3deviation}. 
In the first case (a), we consider the single-atom rate equation model (SARE)~\cite{Ates11}, in which the interactions are absorbed into effective detunings for each atom individually, and no pairs are formed.
The second model (b) accounts for the three possible pairs in the calculation. In addition, each atom in a pair receives an effective detuning due to the interaction with the respective third atom.
In the third model we assume that $V$ is smaller than $V_{12}$, and treat the first two atoms as pair, and the third as individual atom. Again, the mutual interactions between single atom and pair are included in effective detunings. 
In Fig.~\ref{fig:N3deviation} it can be seen that model (a) significantly deviates from the exact treatment at all interaction strengths. Model (b) performs poor at small $V$, because the overlapping pairs mix the exact and approximate interaction between two atoms. However, (b) is best at high interaction strength. The third ansatz (c) reproduces the exact steady state in the limit of $V\rightarrow 0$ by construction, but fails at large $V$, as then correlation with atom 3 cannot be neglected.
Based on this observation, we construct a model in which the two approaches (b) and (c) are combined, depending on the mutual interactions of the atoms. For this, we introduce a critical interaction (or equivalent: a critical distance) that defines when which model has to be used. This is indicated by the dashed red arrow in Fig.~\ref{fig:N3deviation}. The resulting combination of two approaches provides best results over the whole interaction strength range. The increased complexity of our model due to the inclusion of exact two-atom correlations makes the computation based on established techniques impractically slow. We overcome this by a new method discussed below. Since this approach involves a combination of single-atom and pair descriptions, we call it the hybrid model (HM).

\subsection{Derivation of the hybrid model}

We now proceed with the formal analysis. 
For a given ensemble of $N$ atoms, we introduce a critical distance $L_{Cr}$ and use it to divide the set of atoms into pairs and single atoms, as schematically shown in Fig.~\ref{fig:level_scheme}. $L_{Cr}$ should be chosen such that many pairs are included to better incorporate two-atom correlations. But on the other hand, overlapping pairs should be avoided as seen in the $N=3$ case in Fig.~\ref{fig:N3deviation}. Therefore, we use a value which is on the order of the most probable nearest-neighbor distance.
The single-particle Hamiltonian in rotating wave approximation reads
\begin{equation}
	H^{(i)} = \Omega_{12} (S_{12}^{(i)}+S_{21}^{(i)})  +   \Omega_{23} (S_{23}^{(i)}+S_{32}^{(i)}) - \Delta S_{33}^{(i)}\,.
	\label{eqn:H_single}
\end{equation}
Here $S_{ab}^{(i)}$ denotes the operator $\ket{a}_i \, {}_i\bra{b}$ for the \textit{i}th atom and the Rabi frequencies are assumed to be real.
Our hybrid model is based on the full two-body Hamiltonian
\begin{equation}
	H^{(i,j)} = H^{(i)} + H^{(j)} + V_{ij}\; S_{33}^{(i)}\; S_{33}^{(j)}\,,
\end{equation}
with coupling $V_{ij} = C_6 / R_{ij}^6$ \cite{Comparat10,Saffman10}. Including spontaneous emission $\gamma$ as well as dephasing $\Gamma$ described by the standard Lindblad operator $\mathcal L$, the two-atom master equation reads
\begin{align}
	\dot\rho^{(i,j)} = -i \left[ H^{(i,j)},\rho^{(i,j)} \right] + \mathcal{L}[\rho^{(i,j)}]\,.
	\label{eqn:master_N2}
\end{align}
\subsection{Monte Carlo algorithm}

In a first attempt, we generalized the SARE approach~\cite{Ates07,Ates11} to solve Eq.~(\ref{eqn:master_N2}). But it turned out that the calculation of the rates along this algorithm is computationally too inefficient in the two-atom system. Also, negative rates frequently occur, prohibiting a direct Monte Carlo based solving technique. 
The latter problem can be overcome by correcting the calculated transition rates to become strictly positive~\cite{Ates11}. This eliminates the knowledge of the physical time evolution, but still evolves the system into the correct many-body stationary state. 

\subsubsection{Optimizing the rate equations}
Here, we go one step further, and exploit the freedom in choosing the Monte Carlo algorithm (gained by abandoning the physical time evolution) in order to optimize the computational efficiency. For this, we replace the time-consuming calculation of the rates by an evolution equation based on the readily available single-atom steady state populations $\vec{\sigma}^{(\text{SS})} = (\sigma^{(\text{SS})}_1, \sigma^{(\text{SS})}_2, \sigma^{(\text{SS})}_3)^T$ as
\begin{align}
	\label{eq:evolution}
	\frac{d}{dt} \vec{\rho} &= B \vec{\rho}\,,
\end{align}
with vector $\vec{\rho} = (\rho_{11}, \rho_{22}, \rho_{33})^T$ composed of the diagonal elements of the single atom density matrix and 
\begin{align}
B &= \begin{pmatrix} \sigma_{1}^{(\text{SS})} -1 & \sigma_{1}^{(\text{SS})} & \sigma_{1}^{(\text{SS})} \\ \sigma_{2}^{(\text{SS})} & \sigma_{2}^{(\text{SS})} -1 & \sigma_{2}^{(\text{SS})} \\ \sigma_{3}^{(\text{SS})} & \sigma_{3}^{(\text{SS})} & \sigma_{3}^{(\text{SS})} -1 \end{pmatrix}\,. 
\end{align}
In component form, Eq.~(\ref{eq:evolution}) becomes
\begin{align}
\dot{\rho}_{aa} = \sum_{b} \left ( \sigma_b^{(\text{SS})} - \delta_{ab}\right ) \rho_{bb}\,,
\end{align}
with the Kronecker delta $\delta_{ab}$ which is $1$ for $a=b$ and $0$ otherwise.

The matrix $B$ is constructed such that it preserves the total probability,
\begin{align}
\frac{d}{dt}\sum_{i=1}^3 \rho_{ii} = 0\,,
\end{align} and evolves the system into the steady state, as 
\begin{align}
\frac{d}{dt} \vec{\sigma}^\text{(SS)} = B \vec{\sigma}^\text{(SS)} = 0\,.
\end{align}
The tailored propagation matrix $B$ is optimized in the sense that it leads to the same results, but can be calculated much more rapidly than the usual transitions rates obtained from adiabatic elimination of the coherences, as it only depends on the single atom steady state populations $\vec{\sigma}^{\text{(SS)}}$ which are readily available in each Monte Carlo step.

The straightforward extension of this approach for the two-atom populations $\rho_{aa,bb} = \langle S_{aa}^{(i)} S_{bb}^{(j)} \rangle$ of atoms $i$ and $j$ can be written in component form as
\begin{equation}
	\dot \rho_{aa,bb} = \sum_{c,d} \left( \sigma_{cd}^{\text{(SS)}} - \delta_{ac}\delta_{bd} \right) \rho_{cc,dd}\,,
\end{equation}
where $a,b,c,d\in \{1,2,3\}$ denote the atomic states of the two atoms and $\sigma_{cd}^{\text{(SS)}}$ is the steady state of two atoms in states $c$ and $d$.

\subsubsection{Optimizing the Monte Carlo algorithm}
 
Since the optimized evolution equation does not predict the physical intermediate time evolution, we drop all time occurrence in our equations. This allows us to iteratively solve for the many-body steady state with the following Monte Carlo procedure based on the Random Selection Method~\cite{Chotia08,dpa}.
\begin{figure}
	\centering
	\includegraphics{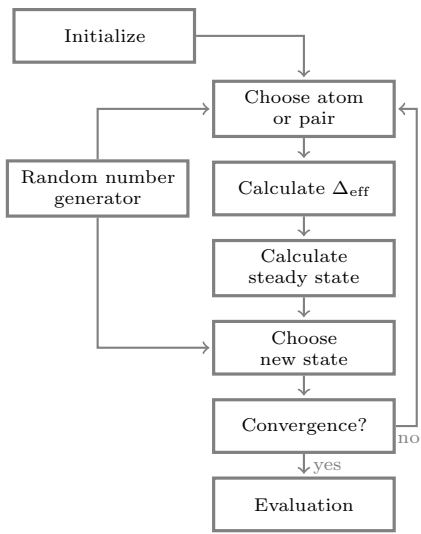}
	\caption{Schematic algorithm of the Monte Carlo method}
	\label{fig:flowchart}
\end{figure}
In Fig.~\ref{fig:flowchart} the algorithm is shown schematically. In the following we describe it in detail.

{\it (i) Choice of atom/pair.} In the first step of the Monte Carlo sequence, the atom or pair is determined whose state will be updated in this step. For this, a random integer is drawn which determines the atom or pair which is handled in the current Monte Carlo step. In the following, the indices labeling the atoms are denoted as $i$ in case of a single atom and $i_1$ and $i_2$ in case of a pair.

{\it (ii) Effective potential.} Next, the effect of all other atoms on the atom/pair chosen in step (i) is calculated. For this, the interaction potential of the atom/pair with all other atoms is summed up and included in an effective detuning. For a single atom $i$ it reads 
\begin{align}
\Delta_{\text{eff}}^{(i)} = \Delta - {\sum_j}^\prime V_{ij}\,,
\end{align}
with the sum $\sum_j^\prime$ running over all other atoms $j$ which are in the Rydberg state $\ket{3}$. In case of a pair two effective detunings
\begin{subequations}
\begin{align}
\Delta_{\text{eff}}^{(i_1)} = \Delta - {\sum_{j\neq i_2}}^\prime V_{i_1 j}\,,\\
\Delta_{\text{eff}}^{(i_2)} = \Delta - {\sum_{j\neq i_1}}^\prime V_{i_2 j}\,,
\end{align}
\end{subequations}
have to be calculated. Note that the interaction $V_{i_1 i_2}$ is included exactly in the two-atom description of the pair chosen in step (i).

{\it (iii) Steady state.} Next, the steady state of the atom/pair chosen in step (i) under the action of the effective potential calculated in step (ii) is determined. For the single atom $i$, the steady state for the population $\rho_{aa} = \langle S_{aa}^{(i)} \rangle$ is denoted by $\sigma_{a}^\text{(SS)}$. For a pair of atoms $i_1$ and $i_2$, the steady state of $\rho_{aa,bb} = \langle S_{aa}^{(i_1)} S_{bb}^{(i_2)} \rangle$ is labeled as $\sigma_{ab}^\text{(SS)}$. Here, $a,b\in\{1,2,3\}$. 
This step is the most time consuming part of the simulation and should therefore be highly optimized. By comparing different approaches, we found the following method to be the fastest numerically stable approach. The master equation is linear and can be written as $\dot{\vec\rho} = M \cdot \vec\rho$ where $M$ is a sparse matrix of the size $9\!\times\!9$ or $81\!\times\!81$ for single atoms or pairs, respectively. The fact that the master equation is Hermitian can be exploited to obtain a real matrix $M$. The problem of finding the steady state is equivalent to find the eigenvector of $M$ with eigenvalue~$0$. Since all states are coupled, the dimension of the corresponding eigensystem is one and the steady state is unique. $M$ is now decomposed into a product of an orthogonal and an upper triangular matrix in a $QR$ decomposition~\cite{NR92}. Since the matrix $M$ is sparse, Givens Rotations are the best approach to transform $M$ successively into an upper triangular matrix $R$~\cite{NR92}. $R$ does not have 
full rank and the last element on the diagonal vanishes. The steady state can then be obtained by back substitution. In practice, it is useful to arrange the elements of the vector $\vec\rho$ such that the populations are located at the end, because then the back substitution can be stopped earlier.

{\it (iv) State update.} Next, the state of the atom/pair determined in step (i) is modified according to the population probabilities determined in step (iii). For this, a random real number $r$ between 0 and 1 is drawn. In the case of a single atom, the state is changed to $\ket l$, where $l$ is the largest integer with $\sum_{k=1}^{l-1} \sigma_{k}^{\text{(SS)}} < r$. For pairs, the procedure is analogous. Note that the new state can be identical to the old state.

{\it (v) Loop of the Monte Carlo sequence.} The steps (i-iv) are now repeated until the system is converged. We have found that for the parameters considered here, approximately $10\cdot N$ steps are sufficient to ensure convergence.

{\it (vi) Calculation of observables.} Finally, the observables can be evaluated. Typically, an average over many spatial realizations and Monte Carlo trajectories is necessary to obtain good statistics and smooth results expected from larger ensembles of Rydberg atoms. Note that throughout the Monte Carlo evolution, each atom is in a definite atomic state at all times, and the diagonal elements of the density matrix only originate from the ensemble averaging.

With our method, several hundred atoms can be simulated easily. Fitting the runtime up to $N=1000$ results in the scaling law $T\sim N^{1.08}$ and a single realization including 1000 atoms takes about 0.5 seconds on a $3.40$~GHz CPU.
\section{Observables and Comparision with other models}
\begin{figure*}
	\centering
	\includegraphics[width=1.9\columnwidth]{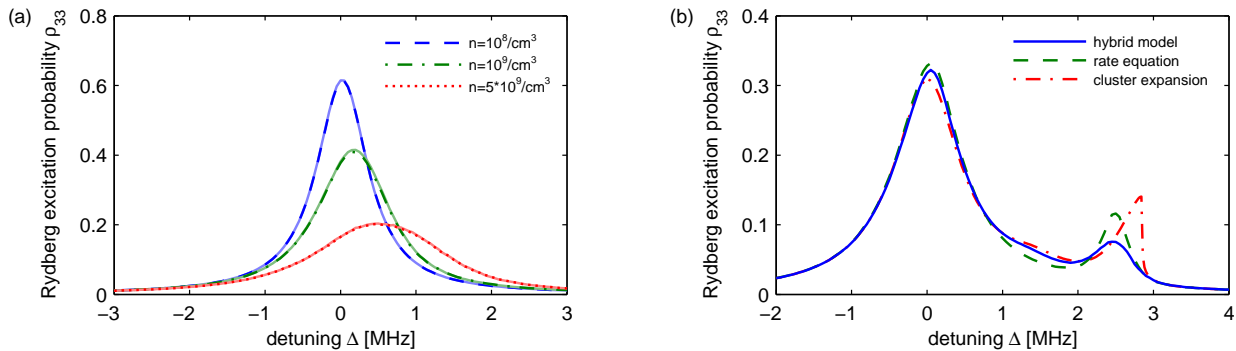}
	\caption{(Color online) \textbf{(a)} Rydberg excitation for a 3D sample of 500 atoms and $C_6 = 50000~\mu\text{m}^6\,\text{MHz}$. The hybrid model (dashed lines) and the single atom rate equation (solid lines) yield the same results. \textbf{(b)} Rydberg excitation on a 1D lattice for different models with interaction $V_{\text{NN}} = 2.5~\text{MHz}$ between adjacent atoms. The resonances at $\Delta=V_{\text{NN}}$ and $\Delta=V_{\text{NN}}/2$ correspond to single-atom and two-atom excitations. Other parameters are $\Omega_{12} = 2~\text{MHz}, \Omega_{23} = 1~\text{MHz}, \gamma_{21}=6~\text{MHz},\gamma_{32}=25~\text{kHz}, \Gamma_{32} = \Gamma_{21} = 100~\text{kHz}$.}
	\label{fig:observables}
\end{figure*}
%
\subsection{Rydberg excitation probability}
In order to verify the validity of the hybrid model we first considered the limit of no pairs, and found results identical to the SARE calculations in~\cite{Ates11}. Next, we calculated the Rydberg population probability 
\begin{align}
\rho_{33} = \frac{1}{N}\sum_{i} \rho_{33}^{(i)}\,,
\end{align}
 and compared the data to established theoretical approaches. In Fig.~\ref{fig:observables}(a), we show results for 3D random samples of atoms with different gas densities. The suppression of excitation at higher densities can be well understood in terms of the Rydberg blockade. In all cases, our hybrid model agrees perfectly with the SARE~\cite{Ates11}, indicating the consistency of our method.
However, significant deviations occur if the geometry is changed to a 1D lattice as shown in Fig.~\ref{fig:level_scheme}(b). Nearest neighbors interact with $V_{\text{NN}} = 2.5$~MHz. The results in Fig.~\ref{fig:observables}(b) show that next to the single-atom excitation peak at $\Delta=0$, the hybrid model predicts a resonance at $\Delta = V_{\text{NN}}$, and an additional weaker resonance at $\Delta = V_{\text{NN}}/2$. 
The first condition describes a resonant excitation process for an atom whose neighbor is already excited to the Rydberg state $\ket 3$, because the laser detuning compensates the interaction energy shift at this point. The second and smaller resonance arises from a 2-photon-process $\ket{22}\rightarrow\ket{33}$ simultaneously exciting two neighboring atoms.
The SARE is not capable of producing this second feature because it does not include the exact two-body interaction. To confirm the presence of the 2-photon-resonance we also performed the simulation with the cluster expansion model (CE) used in~\cite{Schempp10}, which also shows this structure. But interestingly, all three approaches significantly deviate already at $\Delta=V_{\text{NN}}$. It should be noted, however, that the lattice geometry is a significant challenge for all models because of overlapping pairs and non-negligible higher-order correlations.

\subsection{Pair correlation}\label{Sec:pair_correlation}
\begin{figure*}
	\centering
	\includegraphics{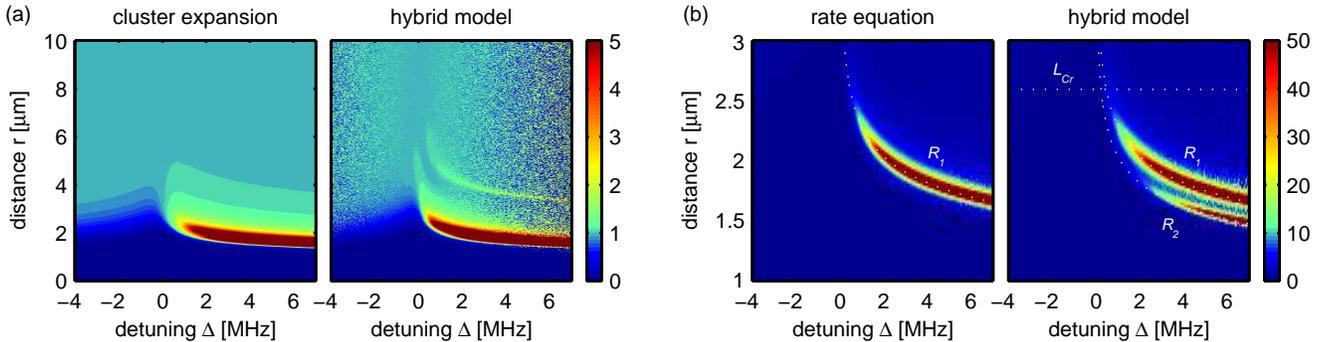}
	\caption{(Color online) Pair correlation function $g^{(2)}(r)$ for different laser detunings. \textbf{(a)} The models reproduce the Rydberg blockade and the uncorrelated regime. In the intermediate region resonances which indicate an excitation enhancement occur. The CE does not yield the line at $2\cdot R_1$. The hybrid model data was obtained from averaging 200000 Monte Carlo runs, noise is due to statistics. \textbf{(b)} The strong resonance lines are shown in detail and the process $\ket{23}\rightarrow\ket{33}$ at $R_1$ can be seen. The hybrid model shows $\ket{22}\rightarrow\ket{33}$ at $R_2$ in addition. Parameters like in Fig.~\ref{fig:observables}, except $C_6=900/2\pi~\mu\text{m}^6~\text{MHz}, n_{\text{1D}} = 0.1~\mu\text{m}^{-1}$. }
	\label{fig:g2}
\end{figure*}
We now turn to our main results on the predictive power of the different models for correlated many-body systems. For this, we calculate the pair correlation function $g^{(2)}(r)$ for a disordered one-dimensional gas. It describes the conditioned probability of having two Rydberg excitations of atoms with distance $r$~\cite{Rob05,wuester10,guenth12,Gaerttner12,Gaerttner12b,Breyel12,Ates12,Stanojevic10}.
We use the definition~\cite{Rob05}:
\begin{equation}
	g^{(2)}(r) = \frac{{\sum_{i,j}}^{(r)} \langle S_{33}^{(i)} S_{33}^{(j)} \rangle}{\rho_{33}^2\cdot {\sum_{i,j}}^{(r)}\,1}\,.
	\label{eqn:g2}
\end{equation}
Here ${\sum_{i,j}}^{(r)}$ denotes the sum over all pairs of atoms with distance $r$. For two uncorrelated atoms the pair correlation function is one. Inside the Rydberg blockade regime, where only one excitation can be present, it becomes zero.

Results for $g^{(2)}(r)$ for different values of the laser detuning $\Delta$ are shown in Fig.~\ref{fig:g2}. As expected, the blockaded and the uncorrelated regimes appear clearly for small and large distances in Fig.~\ref{fig:g2}(a). In between, values $g^{(2)} > 1$ occur for some distances and detunings. These values indicate spatial order, originating from a high probability for multiple excitation at selected distances. Most prominent is the resonance line at about $2~\mu$m which can be characterized by the condition $R_1 = (C_6/\Delta)^{1/6}$. Just as in the lattice simulations, it corresponds to the resonant excitation of a second atom when an atom with distance $R_1$ is already in the Rydberg state. In this case the effective detuning $\Delta_{\text{eff}} = \Delta - C_6/R_1^6$ vanishes. Comparing the results for the hybrid model and CE in Fig.~\ref{fig:g2}(a), it can be seen that the CE does not predict the second peak visible around distances $4~\mu$m and for positive detunings in the HM results. 
This peak occurs if two atoms with distance $R_1$ are already excited, and a third atom again with distance $R_1$ to either of the two atoms is excited in addition. The two outermost atoms in this arrangement then have distance $2\,R_1$. Note that in a higher dimensional geometry this resonance line smears out, because then the distance between the first and the third atom is not necessarily $2\,R_1$. The CE does not predict this higher-order spatial correlation since triply excited states are not part of the model. In contrast, the SARE method correctly predicts the correlations at $2\,R_1$.

Next, we analyze the resonance originating from ordering between nearest neighbors, and show a magnified section of Fig.~\ref{fig:g2}(a) in (b). The left panel of (b) shows the results obtained from the SARE, and contains the single resonance at $R_1$. But in the result of the hybrid model, another resonance appears. Its condition can be determined as $R_2 = [ C_6/(2\Delta) ]^{1/6} \approx 0.89\cdot R_1$. This resonance again originates from the 2-photon-process $\ket{22}\rightarrow\ket{33}$. Interestingly, in this case, the CE correctly predicts the double resonance, whereas the SARE does not. We note that a direct integration of the many-body Schr\"odinger equation describing the many-body Hamiltonian dynamics in two-level systems also shows, and thus confirms, the discussed resonance lines~\cite{Gaerttner12}.

\subsection{Counting statistics}
\begin{figure}
	\centering
	\includegraphics{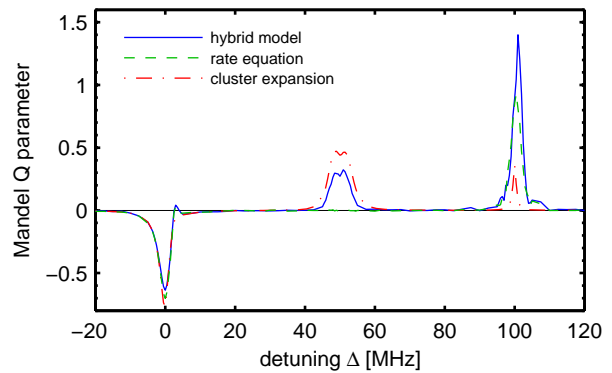}
	\caption{(Color online) Mandel $Q$ parameter for a 1D lattice with 50 atoms. The interaction between adjacent atoms is $V =100$ MHz. Next to the expected sub-poissonian behavior at $\Delta = 0$ we find super-poissonian statistics at $\Delta=V$ and $\Delta = V/2$. Other parameters are $\Omega_{12} = 3~\text{MHz}, \Omega_{23} = 3~\text{MHz}, \gamma_{21}=6~\text{MHz},\gamma_{32}=25~\text{kHz}$.}
	\label{fig:MandelQ}
\end{figure}

Since the hybrid model in principle yields all diagonal elements of the $N$-atom density matrix, it is possible to calculate the full histogram of the number of Rydberg excitations $N_R$. With its first and second moment we obtain the Mandel $Q$ parameter \cite{Mandel79}
\begin{equation}
	Q = \frac{\langle N_R^2 \rangle - \langle N_R \rangle^2}{\langle N_R \rangle} - 1\,.
\end{equation}
This quantity characterizes the counting statistics of the Rydberg excitations. For vanishing detuning $\Delta$ it was found that $Q<0$ \cite{cubellie05,reinh08}, indicating a sub-poissonian behavior. In our simulations for a random gas, this was the case as well, but a more interesting situation arises when we considered an 1D lattice and a varying detuning $\Delta$. In Fig.~\ref{fig:MandelQ} we show the results of the different models.
Next to the well-known sub-poissonian statistics at $\Delta=0$ \cite{Saffman10} we see that the system behaves strongly super-poissonic if the same conditions as for the resonances in the pair correlation function are met, i.e.~when the interaction shift is canceled by the laser detuning. This can be easily understood by recalling that the Mandel $Q$ parameter can be expressed in terms of $g^{(2)}$ \cite{wuester10}. On a 1D lattice this relation reduces to a linear combination of values of $g^{(2)}$, evaluated at multiples of the lattice constant. Consequently, if two adjacent atoms can be resonantly excited, the pair correlation function as well as the Mandel $Q$ parameter increase. In Fig.~\ref{fig:MandelQ} we see that the hybrid model and the cluster expansion cover the two-photon resonance, while the SARE does not include this effect. Again, we note that simulations with the considered models on a lattice are at the border of their range of validity and will in general not give quantitatively exact results.
\subsection{High atom density case}
So far we discussed the hybrid model and its capabilities for low to medium densities where only few atoms are located inside one blockade shell. Since in principle the SARE gives consistent results at also rather high densities \cite{Ates11}, it is of interest if we can also see signatures of the hybrid model's pair treatment in this range. However, if the concept of a maximal $L_{Cr}$ to which exact pair correlations are taken into account is used, the requirement of rarely overlapping pairs diminishes this distance with increasing density. Eventually, $L_{Cr}$ will be much smaller than the blockade radius. In this case, any pair of two atoms characterized in the hybrid approach is perfectly blockaded and does not induce further correlation signatures like a two-photon resonance. Then, the hybrid model described so far does not lead to additional effects compared to the SARE simulations.

To overcome this limitation, we revise the condition when two atoms should be treated as a pair by introducing in addition a limiting lower distance bound $L_{<}$. In the pair correlation function (Sec.~\ref{Sec:pair_correlation}) we have seen 
that for non-vanishing detuning $\Delta$ the main difference of the SARE and the hybrid model is the two-photon resonance at $R_2 = [ C_6/(2\Delta) ]^{1/6}$. This motivates the choice
\begin{align}
L_{<} < R_2 < L_{Cr}
\end{align}
for the boundaries such that only a few overlapping pairs are included in the simulation. Indeed, our simulations confirm, that this procedure allows to identify signatures of two-atom correlations also at higher densities. 

To determine the accuracy of the results from the hybrid model in the high density regime, a reference is required. Next to experimental data, the only presently available choice are exact solutions obtained from time integrating Schr\"odinger's equation for the many-body system in a truncated Hilbert space. However, such Hilbert space truncations are based on the Rydberg blockade, which allows to restrict the total Hilbert space to those states with few excitations, thereby circumventing the exponential growth of the state space with the particle number. But this approach cannot be applied to three-level systems as studied here, as the intermediate (non-Rydberg) excited state is not affected by interactions between the atoms. Therefore, excitations of this intermediate excited state are not restricted, leading again to an exponential scaling of the required state space with the atom number. Thus, we can only use a cloud of two-level systems with a ground and a Rydberg state for the following comparison. In particular, we focus on the observable
\begin{align}
\label{obs}
\frac{{\sum_{i,j}}^{(r)} \langle S_{33}^{(i)} S_{33}^{(j)} \rangle} {{
\sum_{i,j}}^{(r)}\,1}\,,
\end{align}
which is the probability that two atoms with distance $r$ are excited simultaneously. From comparison of Eq.~(\ref{obs}) with the pair correlation function $g^{(2)}$ in Eq.~(\ref{eqn:g2}), and noting that $g^{(2)}(r)\to 1$ for large distances $r$, we find that at large $r$, Eq.~(\ref{obs}) characterizes the total excitation probability squared $\rho_{33}^2$. 

\begin{figure}
	\centering
	\includegraphics{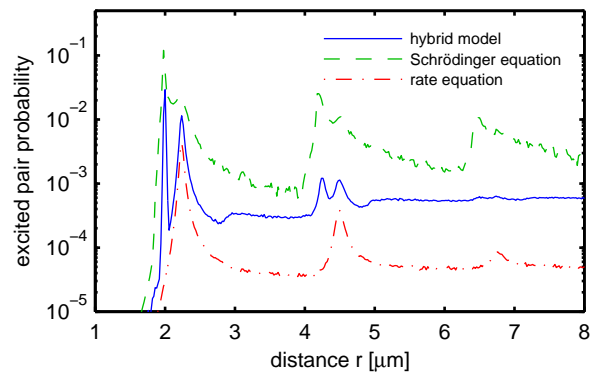}
	\caption{(Color online) Comparison of computational approaches in the high density case. The hybrid model is extended to use both a lower and an upper bound to include pairs as explained in the text. Parameters: $\Omega=1~\text{MHz}, \Delta=7~\text{MHz}, C_6 = 900~\mu\text{m}^6~\text{MHz}, n_{\text{1D}} = 3~\mu\text{m}^{-1}$.}
	\label{fig:HighDensG2}
\end{figure}
Results for the different models are shown in Fig.~\ref{fig:HighDensG2}. Clearly, the hybrid model and the SARE both underestimate the higher order correlations visible in the exact solution. This is expected, since higher-order correlations are not fully included in these models. But over the entire distance range, the HM result is much closer to the full solution than the SARE result. Also, in contrast to the SARE, the hybrid model is capable of capturing the two-photon resonance, leading to the double peak structure close to the blockade radius at distance $\approx 2\,\mu$m. Furthermore, the result at large distance $r$, which is approximately the excitation probability squared $\rho_{33}^2$, is severely underestimated by the SARE, while the HM yields more consistent results.
The shape of the peak structure in the Schr\"odinger equation, which directly depends on the interaction potential \cite{Gaerttner12b}, differs from the HM results. Since the HM covers pairs only up to $L_{Cr}$, the long tail of the two-photon resonance towards larger $r$ cannot be reproduced. Nevertheless, already the narrow range $L_{<} \le r \le L_{Cr}$ provides significant improvements.

We want to emphasize that the hybrid model included 1000 atoms in this calculation, while the Schr\"odinger equation simulation was performed with 45 atoms. Therefore, finite size effects overestimating the population probability and the peak heights might have a small influence on the system \cite{Gaerttner12}. To increase the trap length and the number of atoms in the Schr\"odinger case is computationally unfeasible, which demonstrates that rate equation based models generally cover complementary application ranges in terms of atom numbers.
\begin{figure}
	\centering
	\includegraphics{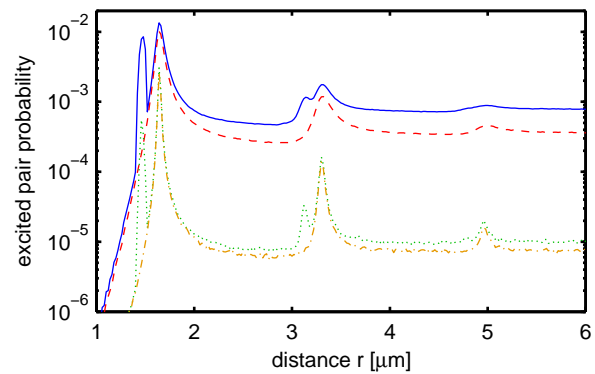}
	\caption{(Color online) Probability of excited pairs for three-level atoms in the high density regime. Simulation results for HM and SARE with 1000 atoms are shown for parameters of Fig.~\ref{fig:MandelQ} (solid and dashed) and Fig.~\ref{fig:observables} (dotted and dash-dotted), except $\Delta=7~\text{MHz}, n_{\text{1D}} = 3~\mu\text{m}^{-1}$.}
	\label{fig:HighDensG2_3States}
\end{figure}

Going beyond the capabilities of numerical integrations of the Schr\"odinger equation on restricted Hilbert spaces, we also performed simulations for three-level systems in the high-density regime. Results for both the HM and the SARE are shown in Fig.~\ref{fig:HighDensG2_3States}. As in the two-level case, the two methods lead to different predictions for the total excitation probability, which can be obtained from the asymptotic values for large $r$. Also, the double peak structures of the resonances are only recovered in the Hybrid model, as in the low-density case. 

It should be noted that a priori, there is no formal justification for the application of the models discussed here in the high density case. These models originate from a few-atom description and therefore cannot take into account the full correlations of the many-body system. Furthermore, based on our present analysis, it is not possible to estimate a parameter range over which the rate equation based models are meaningful approximations in the high density regime. But nevertheless, we can conclude from our data that the HM has consistently significant advantages over the SARE approach also in the high density regime.

Establishing a rate equation approach in the high density regime would be highly desirable, as they would enable one to model multi-level atoms with intermediate non-Rydberg state as used in many recent experiments, in contrast to numerical simulations on truncated Hilbert spaces. Also, typically larger number of atoms can be treated in rate equation based models than in simulations on truncated Hilbert spaces.
\section{Conclusions}
In summary, we presented a hybrid model for the simulation of large ensembles of Rydberg atoms. It is based on an exact two-atom calculation for pairs with distances below a characteristic length scale $L_{Cr}$, combined with an approximate treatment via effective detunings for more distant atoms. We proposed a method to iteratively solve for the steady state of the hybrid model based on the Random Selection Method without the need for time consuming calculations of transition rates. This way, high predictive power is combined with fast calculation times. We found agreement of the hybrid approach with existing models for simple observables such as the Rydberg excitation probability. But different predictions are found for pair correlation function $g^{(2)}(r)$. The HM predicts structures which both, the CE and the SARE models, can only partially reproduce. We identified the additional structures as originating from multi-atom processes and can confirm them in the Mandel $Q$ parameter. This not only 
demonstrates the capability of the hybrid model to characterize higher-order correlations, but also that higher-order correlations cannot be neglected even at low densities. Finally, we discussed an expansion of the HM to higher densities, which is desirable since in contrast to many-body simulations on truncated Hilbert spaces, rate equation based models allow for the simulation of multi-level systems with intermediate non-Rydberg states.
\section*{Acknowledgment}
Financial support by the Heidelberg Center for Quantum Dynamics and the Helmholtz Association (HA216/EMMI) is gratefully acknowledged.

\end{document}